\documentclass{article} % For LaTeX2e
\usepackage{iclr2017_workshop,times}
\usepackage{hyperref}
\usepackage{url}
\usepackage{amssymb}

\usepackage{graphicx} 
\usepackage{caption}
\usepackage{subcaption}
\usepackage{color}
\usepackage{wrapfig}

\title{Semantic embeddings for program behavior patterns}

\author{Alexander Chistyakov${}^1$, Ekaterina Lobacheva${}^{1,2}$, Arseny Kuznetsov${}^1$, Alexey Romanenko${}^1$ \\ 
${}^1$Detection Methods Analysis Group, Kaspersky Lab \\
${}^2$Faculty of Computer Science, National Research University Higher School of Economics\\
Moscow, Russia\\
\texttt{\{Alexander.Chistyakov, Ekaterina.Lobacheva, Arseny.Kuznetsov, } \\
\texttt{Alexey.Romanenko\}@kaspersky.com} \\
}

\begin{document}

\maketitle

\begin{abstract}

In this paper, we propose a new feature extraction technique for program execution logs. First, we automatically extract complex patterns from a program's behavior graph. Then, we embed these patterns into a continuous space by training an autoencoder. We evaluate the proposed features on a real-world malicious software detection task. We also find that the embedding space captures interpretable structures in the space of pattern parts.

\end{abstract}

\section{Introduction}

Malware, or malicious software, is a key element of cyberattacks that damages companies and individuals worldwide and benefits criminals. Nowadays, malware is concealed using obfuscation, encryption, anti-emulation and other techniques, making detection of malware significantly harder. 
Classifying whether a previously unseen file is a malware or a benign program is an important challenge for cybersecurity companies. Employing machine learning methods for this problem is a promising area of research.

The two broad classes of malware analysis techniques are \textit{static} and \textit{dynamic}. The former's methods operate on raw binary files, leading to significant issues with analysis of encrypted and obfuscated files. The latter's approach consists of executing the binary file in a controlled environment and monitoring its behavior. The dynamic approach is more time- and resource-consuming, but it provides higher accuracy. Behavior monitoring results can typically be presented as a log of the observed system events or API calls (function name, arguments, and, optionally, a return value). Currently, the most popular approach for feature extraction from such logs is to construct a set of different indicator features such as n-grams of events or links between APIs and their arguments (\cite{lsh_log_clustering,invincea_win_audit,nn_on_indicators,fisher_info_indicators}). Some other classification methods apply recurrent neural networks to a sequence of notes in a log (\cite{microsoft_rnn, strange_rnn}). However, the latter approach is sensitive to the mixing of lines in a log caused by multiprocessing, intentional obfuscation by malware, and difference in the execution environments.

In this paper we propose a new feature extraction technique for logs that is based on specific behavior graphs. We consider a graph as a union of behavior patterns (specific subgraphs) and construct a feature representation of a log by combining feature vectors of these patterns. To extract a compact and meaningful continuous feature representation for behavior patterns, we train an autoencoder.
In the experiments, we show that the log representation constructed by our technique provides high classification accuracy on a large
real-world 
dataset. In addition, we illustrate the ability of our model to automatically capture interpretable structure in the space of pattern parts similarly to the word2vec model (\cite{word2vec}).

\section{Feature representation for Logs}

\begin{figure}[h]
    \centering
        \begin{subfigure}[b]{0.24\linewidth}
                \centering
                \includegraphics[width=\textwidth]{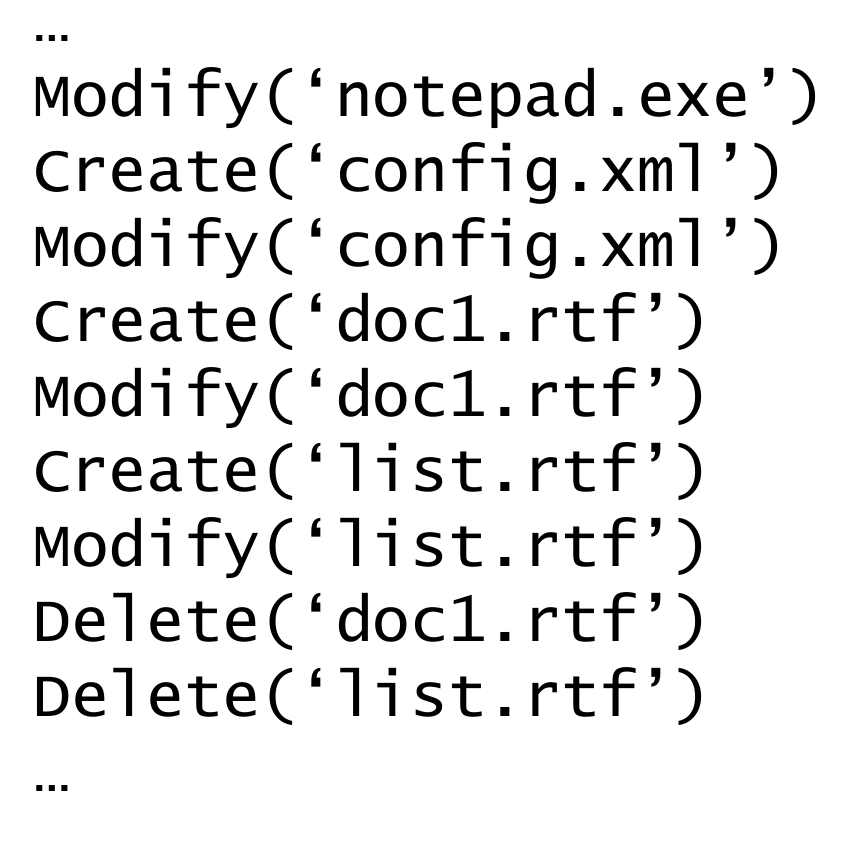}
                \subcaption{Example of a log.\label{fig:log}}
        \end{subfigure}%
        \begin{subfigure}[b]{0.40\textwidth}
                \centering
                \includegraphics[width=\textwidth,page=1]{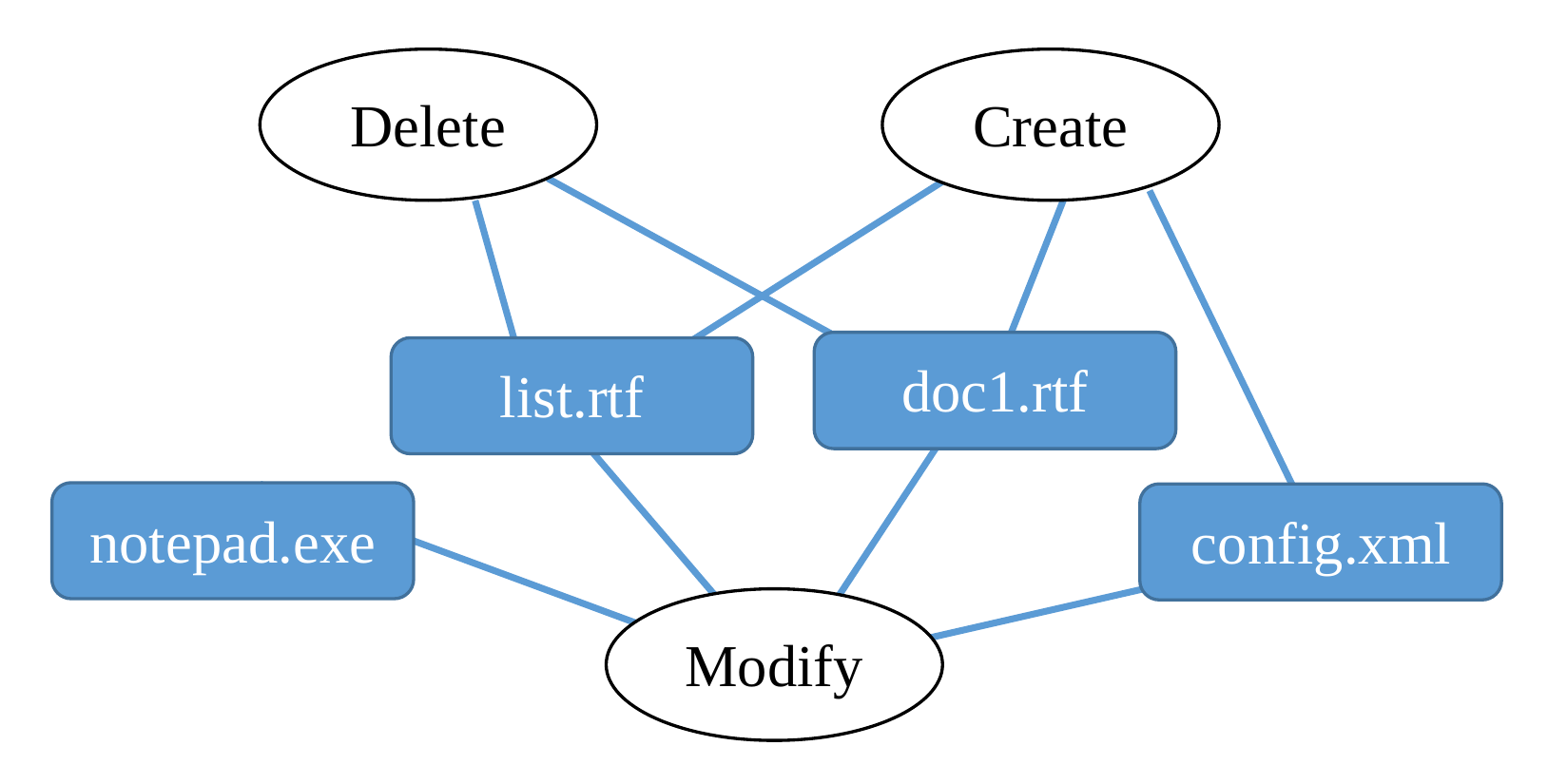}
                \vspace{0.03\linewidth}
                \subcaption{Behavior graph.\label{fig:graph}}
        \end{subfigure}%
        \begin{subfigure}[b]{0.36\textwidth}
                \centering
                \includegraphics[width=\textwidth]{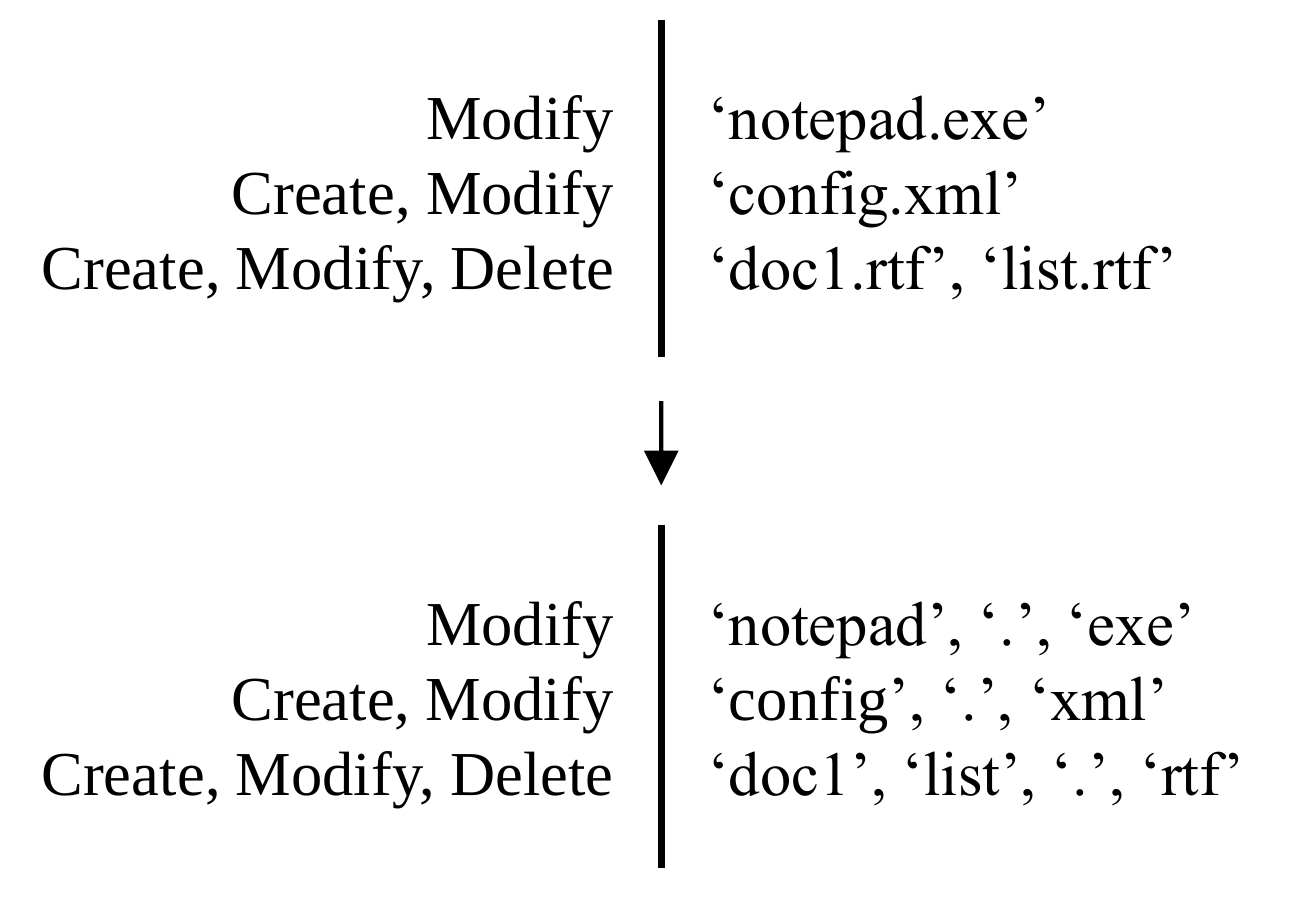}
                \subcaption{Extracted patterns. \label{fig:features}}
        \end{subfigure}
        
        \caption{Graph representation of a log and pattern extraction process.}
        \vspace{-0.4cm}
        \label{fig:DBM}
\end{figure}

In this paper, 
a log means a sequence of all system events that occurred during program execution alongside with their arguments. A toy example of such a log is presented in Figure~\ref{fig:log}. Each line of a log corresponds to one system event and contains the type of the event and one or more arguments (file names, URLs, memory addresses, etc.).

\vspace{-0.3cm}
\paragraph{Graph representation of a log} Because the sequence representation of logs is unstable,
we represent them with behavior graphs. Such representation does not rely on the event order but still captures all event-argument interactions. We define a behavior graph as a bipartite graph whose nodes correspond to event types and arguments occurred in the log. Two nodes 
are connected with an edge
if and only if
the corresponding event type and argument occur together in the same system event in the log. The graph constructed for the toy example log is presented in Figure~\ref{fig:graph}.

Lots of papers on representation learning for graphs have appeared in the last few years (\cite{dgk, node2vec, subgraph2vec}); however, they are all limited by a finite node label space.
In our problem, we deal with a very diverse space of node labels because of the natural variety of the arguments such as file paths, URLs, and so forth.

To construct a feature representation of a log we extract patterns of program behavior from the graph, build an embedding of these patterns into the space $\mathbb{R}^D$
and then combine separate feature vectors into a graph description.

\vspace{-0.3cm}
\paragraph{Behavior pattern extraction} The set of the event types that share the same argument in the graph represents some pattern of the program behavior. Furthermore, it is typically important to capture the fact that a sequence of system events is repeated with different arguments in a log. For example, malware could
continuously modify items in the same
subtree of a file system
or try to connect to many different hosts in a row. Arguments themselves also matter: It is one thing when the program downloads and runs a Windows update
and completely another when it starts
an unknown \textit{*.exe} file from a suspicious server. Therefore we define a behavior pattern as the set of system events and arguments such that all of the events share all of the arguments, and each argument corresponds only to these events and nothing else. 
To extract patterns from the graph, we first find all adjacent event types for each argument and then combine arguments with the same event sets into a single pattern. 

At this stage, the selected patterns may be unique, but their arguments may have a lot in common with some other previously observed arguments.
For example, the particular file name \texttt{C:{\textbackslash}Windows{\textbackslash}374683.ini} may occur only once in the whole dataset, but the disk name \textit{C}, folder name \textit{Windows} and file extension \textit{ini} are very common. Therefore we propose splitting each argument into a set of tokens by separators (such as '://','.',':', etc.). Here, separators are also considered to be tokens because the type of the argument can be determined from them. Pattern extraction process for the toy example log is illustrated in Figure~\ref{fig:features}. 

As a result, each pattern can be represented as a sparse binary vector of length $M+K$, where the first $M$ features correspond to event types and the next $K$ features correspond to tokens. Here $M$ is the number of all different event types that exist in the logs, and it is fixed by the logging system. $K$ denotes the
number of the most frequent tokens observed in the training dataset, and it can be set manually.

\vspace{-0.3cm}
\paragraph{Pattern embeddings} To extract a compact and meaningful feature representation for behavior patterns we train an autoencoder model. The encoded representation $a(x)$ and the reconstruction $\hat{v}(x)$ of a pattern $x$ are obtained from its sparse binary representation $v(x)$ as follows:
\begin{equation}
    a(x) = Wv(x)+b, \quad \phi(x) = \mathrm{ReLU}(a(x)),\quad \hat{v}(x) = \sigma (V \phi(x)+c),
\end{equation}
where $W$ and $V$ are trainable weight matrices of size $(M+K)\times D$ and $D\times (M+K)$ respectively, $b$ and $c$ are trainable bias vectors of length $D$ and $(M+K)$ respectively, $\mathrm{ReLU}(y) = \max(0,y)$ and $\sigma(y) = 1/(1+\exp(-y))$. Then, the reconstruction is compared with the original binary feature vector as follows:
\begin{equation}
    l(x) = -\frac{1}{|P|}\sum_{i\in P}\log(\hat{v}(x)_i) - \frac{1}{|N|}\sum_{i\in N}\log(1-\hat{v}(x)_i),
\end{equation}
where $P$ is the set of
non-zero
elements in $v(x)$, $N$ is a random subset of zero elements in $v(x)$ and $|\cdot|$ is the cardinality of the set. We take only a small subset of zero elements for efficiency, similar to the negative sampling technique (\citet{word2vec}). 
Training proceeds by optimizing the sum of reconstruction cross-entropies across the training set using Adam (\citet{adam}).

To construct a fixed-size feature representation for a log, we combine the encoded feature vectors $a(x)$ of all patterns from this log by applying the element-wise min, max, and mean functions. As a result, for each log, we have a final feature vector of length $3D$.
 
\section{Results/Experiments}

\begin{wrapfigure}[15]{r}{.5\textwidth} 
\vspace{-5ex}
\includegraphics[width=.5\textwidth]{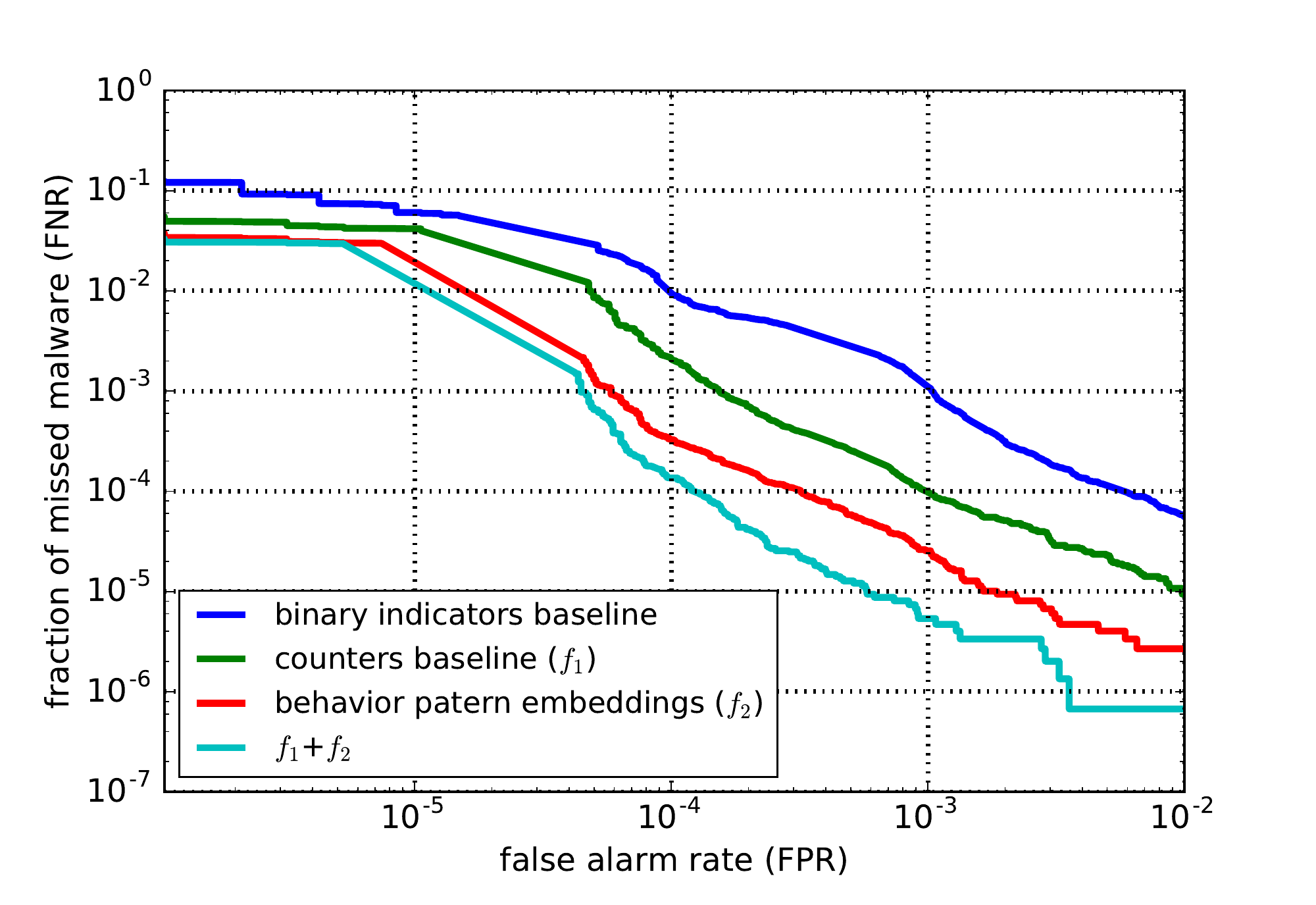}
\caption{Comparison of feature representations of logs ($f_2$ is ours).}
        \label{fig:ROC-AUC}
\end{wrapfigure}

Comparison of different dynamic malware detection methods is difficult because researchers use different sandboxes to collect their data, and none of the datasets are publicly available. To evaluate our log representation, we collected $4\,964\,506$ malicious and $3\,145\,853$ benign logs from our in-lab sandbox and a set of
user-managed
machines, and randomly split our data into train (70\%) and test (30\%) sets. 

The majority of existing articles on dynamic malware detection are based on constructing a set of indicator behavior features,
so we selected the set of the most frequent groups of events that share the same argument
from our training data, and use indicator features constructed from them as a baseline feature representation. We also try to use counters instead of binary indicators.
We train the XGBoost model (\cite{xgboost}) with 300 trees of depth 30 as a classifier to compare our features to baselines. Figure~\ref{fig:ROC-AUC} shows that the usage of our semantic features 
reduces the number of missed malicious samples several times
while keeping an extremely low false alarm rate. 
A combination of our semantic features and baseline counter features provides even better performance.

\begin{table}[!htb]

    %\vspace{-0.1cm}

    %\small
    %\footnotesize
    \scriptsize
    %\tiny

    %\caption{Global caption}
    \begin{minipage}[t]{.5\linewidth}
    % \begin{table}[t]
        \caption{Synonyms locality embedding}
        \label{tab:nearest_neigbors}
        \begin{center}
        %\begin{tabular}{ll}
        \begin{tabular}{r|l}
        %\multicolumn{1}{c}{\bf TOKEN}  &\multicolumn{1}{c}{\bf TOP-5 NEAREST NEIGHBORS}
        {\bf TOKEN}  &{\bf TOP-5 NEAREST NEIGHBOURS}
        \\ \hline \\
        word    &   excel, dotm, outlook, machine, open \\
        com     &   www, ://, http, net, ru \\
        jpg     &   png, gif, css, xml, html \\
        js      &   txt, htm, ie5, css, cookies \\
        
        34      &   36, 42, 56, 38, 35 \\
        3960    &   3964, 3972, 3952, 3968, 3956 
        %3801088 &   5177344, 5570560, 6160384, 3276800, 5963776 \\
    
        \end{tabular}
        \end{center}
    % \end{table}
    \end{minipage}
    \begin{minipage}[t]{.5\linewidth}

    % \begin{table}[t]
        \caption{Arithmetic operations on tokens}
        \label{tab:word2vec_equations}
        \begin{center}
        %\begin{tabular}{ll}
        \begin{tabular}{r|l}
        %\multicolumn{1}{c}{\bf TOKEN}  &\multicolumn{1}{c}{\bf TOP-5 NEAREST NEIGHBORS}
        {\bf TOKEN'S EXPRESSION}  &{\bf RESULT}
        \\ \hline \\
        `word' - `doc' + `xls'  &   `excel' \\
        `video' - `mp4' + `bmp' &   `image'\\
        %`video'-`avi'+`bmp' &   `image'\\  
        
        %`programfiles' - `exe' + `tmp'  &   `appdata'   \\
        `programfiles' - `exe' + `dll'  &   `system32'   \\
        
        `https' - `http' + `80' &   `443'   \\
        `google' - `chrome' + `firefox' &  `mozilla'    \\
        `(' - `)' + `]' &   `['
        \end{tabular}
        \end{center}
    % \end{table}
    \end{minipage}
\end{table}

In order to get some intuition about how our behavior pattern embeddings work, we study the structure of the token embedding space, obtained from rows of the matrix $W$. Table~\ref{tab:nearest_neigbors} shows that tokens that are similar in the vector space are also semantically similar: they form groups of program names, file extensions, and decimal constants.
Another interesting point of the obtained embedding is that it keeps semantic relations between pairs of tokens in the manner of the canonical word2vec model (\cite{word2vec}). Table~\ref{tab:word2vec_equations} presents some of these relations such as the
binding of a file format to its typical
folder or
editor program name, TCP/IP port number to the corresponding web-protocol name, and programming product name to its vendor.

\bibliography{iclr2017_workshop}
\bibliographystyle{iclr2017_workshop}

\end{document}